\documentclass[pra,twocolumn,a4paper,groupedaddress,floatfix,aps,10pt]{revtex4-1}
\usepackage{graphicx,amsmath,amssymb,amsfonts,dsfont}
\newcommand{\mf}[1]{\boldsymbol{#1}}
\newcommand{\ket}[1]{\ensuremath{|#1\rangle}}
\newcommand{\mc}[1]{\ensuremath{\mathcal{#1}}}
\newcommand{\bra}[1]{\ensuremath{\langle #1 |}}

\newcommand{\imag}{\mathrm{i}}
\begin{document}
\title{Topological spin models in Rydberg lattices}
\author{Martin Kiffner${}^{1,2}$}
\author{Edward O'Brien${}^{2}$}
\author{Dieter Jaksch${}^{2,1}$}

\affiliation{Centre for Quantum Technologies, National University of Singapore,
3 Science Drive 2, Singapore 117543${}^1$}
\affiliation{Clarendon Laboratory, University of Oxford, Parks Road, Oxford OX1
3PU, United Kingdom${}^2$}

\begin{abstract}
We show that resonant dipole-dipole interactions between Rydberg atoms in a triangular lattice can give rise 
to artificial magnetic fields for spin excitations. We consider 
the coherent dipole-dipole coupling between $np$ and $ns$ Rydberg states and derive an effective spin-1/2 Hamiltonian for the $np$ excitations. 
By breaking time-reversal symmetry via external fields we engineer complex hopping amplitudes for 
transitions between two rectangular sub-lattices. The phase of these hopping amplitudes depends on the direction of the 
hop. This gives rise to a staggered, artificial magnetic field which induces 
non-trivial topological effects. We calculate the single-particle band structure 
and investigate its Chern numbers as a function of  the lattice parameters and the detuning 
between the two sub-lattices. We identify extended parameter regimes where the  Chern number of 
the lowest band is $C=1$ or $C=2$. 
\end{abstract}
\maketitle
\section{Introduction \label{introduction}}
Regular arrays of ultracold neutral atoms~\cite{jaksch:98,bloch:05} are a versatile tool 
for the quantum simulation~\cite{calarco:00,cirac:12,Johnson:14} of many-body physics~\cite{bloch:08}. 
Recent experimental progress allows one to control and observe atoms with single-site
resolution~\cite{bakr:09,bakr:10,sherson:10,weitenberg:11,gericke:08,wuertz:09}
which makes dynamical phenomena experimentally accessible in these systems. 
One promising perspective is to use this setup for investigating the rich physics of 
quantum magnetism~\cite{auerbach:94,simon:11,sanner:12} and strongly correlated spin systems that are 
extremely challenging to simulate on a classical computer. 
However, the simulation of magnetic phenomena with cold atoms faces two key challenges.
First, neutral atoms do not experience a Lorentz force in an external magnetic field. 
In order to circumvent this problem, tremendous effort
has been made to create artificial gauge fields for neutral 
atoms~\cite{ruseckas:05,dalibard:11,dum:96,lin:09,lin:09n,lin:11,aidelsburger:11,jaksch:03,struck:11,struck:12,jimenez_garcia:12,palmer:06,palmer:08,cooper:13,jotzu:14,hauke:12,zhang:13}.  
For example, artificial magnetic fields allow one to investigate the integer~\cite{sterdyniak:15} and fractional 
quantum Hall effects~\cite{palmer:06,palmer:08,cooper:13} with cold atoms, and the experimental realization of the topological Haldane model was 
achieved in~\cite{jotzu:14}.
Second,  cold atoms typically interact via  weak contact interactions. Spin systems with 
strong and long-range interactions can be achieved by admixing van der Waals interactions between Rydberg states~\cite{glaetzle:14,glaetzle:15} 
or by replacing atoms with dipole-dipole interacting polar molecules~\cite{peter:12,gorshkov:11a,gorshkov:11b}. 
In particular, it has been shown that the dipole-dipole interaction can give rise to topological flat bands~\cite{yao:12,peter:15} and 
fractional Chern insulators~\cite{yao:13}. 
The creation of bands with Chern number $C=2$ via resonant exchange interactions between polar molecules  has been explored in~\cite{peter:15}.

Recently an alternative and very promising platform for the simulation of strongly correlated spin systems has emerged~\cite{barredo:15}.  
Here resonant dipole-dipole interactions between Rydberg atoms~\cite{gallagher:ryd} 
enable quantum simulations of spin systems at completely different length scales compared with polar molecules. 
For example, the experiment in~\cite{barredo:15} demonstrated the realization of the $XY$  Hamiltonian 
for a chain of atoms and with a lattice spacing  of the order of $20\mu\text{m}$. 
At these length scales, light modulators allow one to trap atoms in arbitrary, two-dimensional geometries and 
to apply custom-tailored light shifts at individual sites~\cite{nogrette:14,labuhn:16,barredo:16}. 
The resonant dipole-dipole interaction is also ideally suited for the investigation of transport phenomena~\cite{robicheaux:14b,bettelli:13,schempp:15} 
and can give rise to artificial magnetic fields acting on the relative motion of two Rydberg atoms~\cite{zygelman:12,kiffner:13,kiffner:13b}. 
Here we show how to engineer artificial magnetic fields for spin excitations in two-dimensional arrays of 
dipole-dipole interacting Rydberg atoms. More specifically, we consider a triangular lattice of Rydberg atoms as shown in Fig.~\ref{fig1} 
where the resonant dipole-dipole interaction enables the coherent exchange of excitations between atoms in  $np$ and $ns$ states.
We derive an effective spin-1/2 Hamiltonian for the $np$ excitations with  complex hopping 
amplitudes giving rise to artificial, staggered magnetic fields. This results in non-zero 
Chern numbers of the single-particle band structure, and the value of the Chern number in the lowest band can be  adjusted to $C=1$ or $C=2$ by 
changing the lattice parameters.

Note that in our system all atoms comprising the lattice are excited to a Rydberg state. This is in contrast to 
the work in~\cite{glaetzle:14,glaetzle:15}, where the atoms mostly reside in their ground states and the population in the Rydberg manifold is small. 
Consequently, our approach is in general more vulnerable towards losses through spontaneous emission. On the other hand, 
the magnitude of the resonant dipole-dipole interaction is much stronger compared with a small admixing of van der Waals interactions, 
and hence the coherent dynamics takes place on much shorter time scales. In addition, the distance between the atoms can be much larger in 
our approach which facilitates the preparation and observation of the excitations. 
This paper is organised as follows. We give a detailed description of our system  in Sec.~\ref{model} where we 
engineer an effective  Hamiltonian for the $np$ excitations. 
We then investigate the single-particle band structure and provide a systematic investigation of 
the topological features of these bands as a function of the system parameters in Sec.~\ref{results}.
A brief summary of our work is presented in Sec.~\ref{summary}.
\section{Model \label{model}}
%
\begin{figure}[t!]
\begin{center}
\includegraphics[width=8cm]{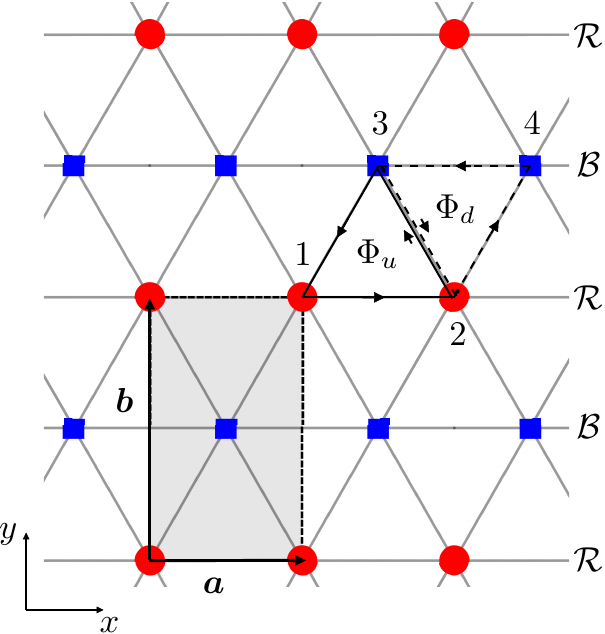}
\end{center}
\caption{\label{fig1}
(Color online) Triangular lattice of Rydberg atoms in the $x-y$ plane. The lattice is comprised of two rectangular sub-lattices 
$\mc{R}$ and $\mc{B}$ that are shifted by $\vec{a}/2+\vec{b}/2$ with respect to each other, where 
$\vec{a} = a \vec{e}_x$ and $\vec{b} = b \vec{e}_y$ are the primitive basis vectors of each sub-lattice.
The sites of the $\mc{B}$ ($\mc{R}$) lattice are indicated by blue squares (red dots). 
The unit cell of the whole lattice is shown by the shaded area and contains two lattice sites.  
$\Phi_u$ is the flux through the upward pointing triangle $1\rightarrow2\rightarrow3\rightarrow1$, and 
$\Phi_d$ is the flux through the downward pointing triangle $2\rightarrow4\rightarrow3\rightarrow2$. 
}
\end{figure}
%
We consider  a two-dimensional triangular lattice of Rydberg atoms in the $x-y$ plane as shown in Fig.~\ref{fig1}. 
Each lattice site contains a single Rydberg atom which we assume to be pinned to the site. The triangular lattice is comprised 
of two rectangular sub-lattices $\mc{B}$ and $\mc{R}$ that are labelled by blue squares and red dots in Fig.~\ref{fig1} respectively. 
Each sub-lattice is described by two orthogonal  primitive basis vectors
$\vec{a} = a \vec{e}_x$ and $\vec{b} = b \vec{e}_y$, and the two sub-lattices are shifted by $\vec{a}/2+\vec{b}/2$ with respect to each other. 
In the following, we derive an effective spin-1/2 model for Rydberg excitations in the $np$ manifold over a background of $ns$ 
states with principal quantum number $n\gg 1$. 
After introducing the general Hamiltonian of the system, we first engineer an effective Hamiltonian for $np$ excitations on the $\mc{B}$ 
sub-lattice. We then apply the same procedure to the $\mc{R}$ sub-lattice but choose a different $np$ state compared to the $\mc{B}$ atoms. 
Finally, we show that the dipole-dipole interaction couples the two sub-lattices and the corresponding Hamiltonian contains 
complex hopping amplitudes giving rise to artificial magnetic fields. 

\begin{figure}[t!]
\begin{center}
\includegraphics[width=7.5cm]{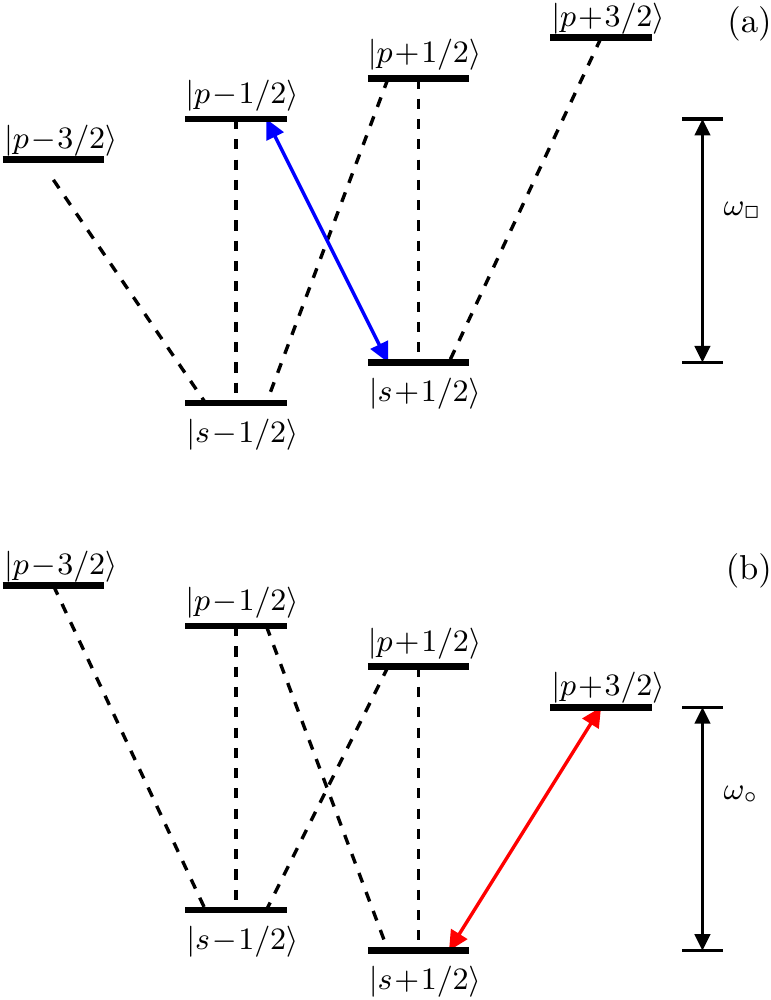}
\end{center}
\caption{\label{fig2}
(Color online) 
The level scheme of each atom consists of the $ns_{1/2}$ and $np_{3/2}$ manifolds. Dashed lines denote allowed dipole transitions. 
(a) The effective spin-1/2 system at  sites $\mc{B}$ is formed by states $\ket{p_{3/2}-1/2}$ and $\ket{s_{1/2}1/2}$. 
The corresponding dipole transition with  transition frequency $\omega_{\Box}$ is indicated in blue. 
(b) The effective spin-1/2 system at  sites $\mc{R}$ is formed by states $\ket{p_{3/2}3/2}$ and $\ket{s_{1/2}1/2}$. 
The associated dipole transition  with transition frequency $\omega_{\circ}$ is indicated in red.
}
\end{figure}

The atomic level scheme of each atom 
is comprised of two angular momentum manifolds $ns_{1/2}$ and $np_{3/2}$ with principal quantum number $n\gg 1$ as shown in
Fig.~\ref{fig2}. The  Zeeman sublevels of each multiplet are denoted by $\ket{l_jm}$, where $l$ labels the orbital angular
momentum,  $j$ is the total angular momentum and the projection of the electron's angular momentum onto the $z$-axis is
denoted by $m$. The Hamiltonian of a single atom at site $\alpha$ is   given by 
\begin{align}
 H_{\alpha}^{(0)} = &  \hbar  \omega_p \sum\limits_{m=-3/2}^{3/2} \ket{ p_{3/2} m }_{\alpha} \bra{ p_{3/2}m }_{\alpha} \notag \\
& + \hat{L}_{\alpha}[p_{3/2}] + \hat{L}_{\alpha}[s_{1/2}] \, ,
 \label{h0}
\end{align}
where the first line is the Hamiltonian for the degenerate $np_{3/2}$ manifold in the absence of external fields, $\hbar\omega_p$ 
is the energy of the $np_{3/2}$ multiplet and we set the frequency of the $ns_{1/2}$ multiplet $\omega_s=0$. 
In the second line of Eq.~(\ref{h0}), $\hat{L}_{\alpha}[l_j]$ are level shift operators removing the Zeeman degeneracy of the multiplet $l_j$ at site $\alpha$. 
An example for the operators $\hat{L}_{\alpha}[l_j]$ is given in Eq.~(\ref{ShiftOp}) at the end of Sec.~\ref{model}. 
In the following  we assume that all atoms in rows labelled by $\mc{B}$ and indicated by a blue square in  Fig.~\ref{fig1} experience the same level shifts. Similarly, 
all atoms in rows labelled by $\mc{R}$ and indicated by a red dot in Fig.~\ref{fig1} have equivalent level schemes. However, atoms in sites $\alpha\in\mc{R}$ 
have a different internal level structure compared with atoms in sites $\alpha\in\mc{B}$. 
The full Hamiltonian for the system shown in Fig.~\ref{fig1} is then given by 
\begin{align}
 H = \sum\limits_{\alpha} H_{\alpha}^{(0)}   + 
 \frac{1}{2}\sum\limits_{\alpha,\beta\atop\alpha\not=\beta} V_{\alpha\beta}\, ,
 \label{fullham}
\end{align}
where $V_{\alpha\beta}$ is the dipole-dipole interaction~\cite{tannoudji:api} between atoms at sites $\alpha$ and $\beta$,
\begin{align}
V_{\alpha\beta} = \frac{1}{4\pi\varepsilon_0 R^3}[\vec{\hat{d}}^{(\alpha)}\cdot\vec{\hat{d}}^{(\beta)}
-3(\vec{\hat{d}}^{(\alpha)}\cdot\tilde{\vec{R}})(\vec{\hat{d}}^{(\beta)}\cdot\tilde{\vec{R}})] \,.
\label{vdd}
\end{align}
Here $\varepsilon_0$ is the dielectric constant,  $\vec{\hat{d}}^{(\alpha)}$ is the electric dipole-moment operator of atom $\alpha$, 
$\vec{R} = \vec{R}_{\alpha}-\vec{R}_{\beta}$ is the relative position of the two atoms located at $\vec{R}_{\alpha}$ and $\vec{R}_{\beta}$, respectively, 
and  $\tilde{\vec{R}}=\vec{R}/R$ is the corresponding  unit vector. In the following we consider only near-resonantly coupled states and neglect all 
matrix elements between two-atom states differing in energy by $\Delta E_{\text{FS}}=\hbar\omega_p$ or more. This is justified if the dipole-dipole coupling strength $V_0$ 
is much smaller than the fine structure interval $\Delta E_{\text{FS}}$, which is the case for the typical parameters 
based on rubidium atoms (see Sec.~\ref{results}).

Next we we focus on the $\mc{B}$ lattice and reduce the level scheme at each site  to a two-level system by a suitable choice of the shift operators in Eq.~(\ref{h0}). 
To this end, we assume that the level shifts break the degeneracy of the Zeeman sublevels as shown in Fig.~\ref{fig2}(a) such that all dipole transitions 
can be addressed individually. In particular, we require that the strength of 
the dipole-dipole coupling between nearest neighbours is much smaller than the  splitting between Zeeman sublevels. 
For all $\mc{B}$ atoms, we choose the states $\ket{p_{3/2}-1/2}$ and $\ket{s_{1/2}1/2}$ as the effective spin-1/2 
system. The dipole matrix element of the  $\ket{p_{3/2}-1/2}\leftrightarrow\ket{s_{1/2}1/2}$ transition with 
transition frequency $\omega_{\Box}$ is  (see Appendix~\ref{appendixA})
\begin{align}
 & \vec{d}_{\mc{B}}=\bra{p_{3/2}-1/2}\vec{\hat{d}}\ket{s_{1/2}1/2}  = \mc{D}\frac{1}{\sqrt{6}}\left(\vec{e}_x + \imag\vec{e}_y\right)\,,
 \label{db}
\end{align}
where $\mc{D}$ is the reduced dipole matrix element of the $s_{1/2}\leftrightarrow p_{3/2}$ transition.  
%
%
%
In an interaction picture with respect to the bare atomic energies, the Hamiltonian $H$ in Eq.~(\ref{fullham}) restricted to all $\mc{B}$ atoms can thus be written as 
\begin{align}
 H_{\mc{B}} =  -\frac{1}{6} \sum\limits_{\alpha \in \mc{B} \atop \beta \in \mc{B}} 
 \mc{C}_{\alpha\beta} \left(S_{\alpha}^{+}S_{\beta}^{-}+S_{\beta}^{+}S_{\alpha}^{-}\right) \,, 
 \label{hb}
 \end{align}
where 
\begin{align}
  \mc{C}_{\alpha\beta} & = \frac{|\mc{D}|^2}{4\pi \varepsilon_0 |\vec{R}_{\alpha}-\vec{R}_{\beta}|^3}  
  \label{coupling}
  \end{align}
describes the coupling strength between two $\mc{B}$ atoms located at $\vec{R}_{\alpha}$ and $\vec{R}_{\beta}$, respectively. 
In the following it will be useful to characterise the strength of the dipole-dipole interaction between two atoms separated by $a$, 
and hence we introduce the parameter
\begin{align}
 V_0 = \frac{|\mc{D}|^2}{4\pi \varepsilon_0 a^3} \,.
 \label{strength}
\end{align}
The raising operator for a spin excitation in Eq.~(\ref{hb}) is defined as 
\begin{align}
 S_{\alpha}^{+} = \ket{p_{3/2}-1/2}\bra{s_{1/2}1/2}\,, \quad \quad\alpha\in\mc{B}\,,
 \label{sb}
\end{align}
and its adjoint is the corresponding lowering operator,  $S_{\alpha}^{-}=[S_{\alpha}^{+}]^{\dagger}$. 
Next we follow a similar procedure within the  $\mc{R}$ lattice. In contrast to $\mc{B}$ atoms, 
We choose  the states $\ket{p_{3/2}3/2}$ and $\ket{s_{1/2}1/2}$ as an effective  spin-1/2 system as shown in Fig.~\ref{fig2}(b). 
We assume that all other transitions within $\mc{R}$ atoms are so far-detuned that the dipole-dipole interaction remains restricted to this subsystem. 
We find that the dipole matrix element of the corresponding transition
$\ket{p_{3/2}3/2}\leftrightarrow\ket{s_{1/2}1/2}$  is (see Appendix~\ref{appendixA})
\begin{align}
 &\vec{d}_{\mc{R}}= \bra{p_{3/2}3/2}\vec{\hat{d}}\ket{s_{1/2}1/2} = -\mc{D}\frac{1}{\sqrt{2}}\left(\vec{e}_x-\imag\vec{e}_y\right)\,.
 \label{dr}
\end{align}
The raising operator of this transition with resonance frequency $\omega_{\circ}$ is defined as 
\begin{align}
 S_{\alpha}^{+} = \ket{p_{3/2}3/2}\bra{s_{1/2}1/2}\,, \quad \quad\alpha\in\mc{R}\,, 
 \label{sr}
\end{align}
and  $S_{\alpha}^{-}=[S_{\alpha}^{+}]^{\dagger}$ is the lowering operator. 
In a rotating frame where $S_{\alpha}^{+}$ oscillates with the frequency $\omega_{\Box}$ of excitations in the $\mc{B}$ lattice, the 
Hamiltonian for excitations in the $\mc{R}$ lattice can be written as 
\begin{align}
 H_{\mc{R}} =  \hbar\Delta\sum\limits_{\alpha \in \mc{R}} S_{\alpha}^{+}S_{\alpha}^{-} 
 -\frac{1}{2} \sum\limits_{\alpha \in \mc{R} \atop \beta \in \mc{R}} 
 \mc{C}_{\alpha\beta} \left(S_{\alpha}^{+}S_{\beta}^{-}+S_{\beta}^{+}S_{\alpha}^{-}\right) \,,
 \label{hr}
 \end{align}
where $\Delta = \omega_{\circ}-\omega_{\Box}$ is the detuning between excitations in the  $\mc{R}$ and $\mc{B}$ lattices 
and $\mc{C}_{\alpha\beta}$ is defined in Eq.~(\ref{coupling}). 
For our given geometry and chosen transitions, we find that the dipole-dipole coupling between the two sub-lattices is different from zero.
If the detuning $\Delta$ between $\mc{B}$ and $\mc{R}$ excitations is smaller than the strength of the dipole-dipole coupling between the two sub-lattices, 
the $np$ excitations can hop between the $\mc{B}$ and $\mc{R}$ sites. 
With the expressions for the dipole matrix elements in Eqs.~(\ref{db}) and~(\ref{dr}), the Hamiltonian governing the coupling between the two sub-lattices is 
given by 
\begin{align}
  H_{\mc{B}\mc{R}} = \frac{\sqrt{3}}{2} \sum\limits_{\alpha \in \mc{R} \atop \beta \in \mc{B}} 
 \mc{C}_{\alpha\beta} \left(e^{-2\imag \phi_{\alpha\beta}} S_{\alpha}^{+}S_{\beta}^{-}+e^{2\imag \phi_{\alpha\beta}} S_{\beta}^{+}S_{\alpha}^{-}\right) \, ,  
 \label{hbr}
\end{align}
where 
\begin{align}
  e^{\imag\phi_{\alpha\beta}} & = \left(\tilde{\vec{R}}_{\alpha}-\tilde{\vec{R}}_{\beta}\right)\cdot \left( \vec{e}_x + \imag\vec{e}_y \right) \,.
 \label{phasedef}
\end{align}
Note that the phase $\phi_{\alpha\beta}$ of  excitation hopping between sites $\alpha\in\mc{R}$ and $\beta\in\mc{B}$ is determined by the azimuthal angle  
of the relative position vector $\tilde{\vec{R}}_{\alpha}-\tilde{\vec{R}}_{\beta}$ between the two sites. 

In summary, by restricting the effective level scheme on each site to a two-level system  we obtain 
\begin{align}
 H_{\text{eff}} & =   H_{\mc{B}} +  H_{\mc{R}} +H_{\mc{B}\mc{R}}  \label{effham} \\
 & =  \hbar\Delta\sum\limits_{\alpha \in \mc{R}} S_{\alpha}^{+}S_{\alpha}^{-}  
 -\frac{1}{6} \sum\limits_{\alpha \in \mc{B} \atop \beta \in \mc{B}} 
 \mc{C}_{\alpha\beta} \left(S_{\alpha}^{+}S_{\beta}^{-}+S_{\beta}^{+}S_{\alpha}^{-}\right) \notag\\
 & -\frac{1}{2} \sum\limits_{\alpha \in \mc{R} \atop \beta \in \mc{R}} 
 \mc{C}_{\alpha\beta} \left(S_{\alpha}^{+}S_{\beta}^{-}+S_{\beta}^{+}S_{\alpha}^{-}\right) \notag\\
 &+\frac{\sqrt{3}}{2} \sum\limits_{\alpha \in \mc{R} \atop \beta \in \mc{B}} 
 \mc{C}_{\alpha\beta} \left(e^{-2\imag \phi_{\alpha\beta}} S_{\alpha}^{+}S_{\beta}^{-}+e^{2\imag \phi_{\alpha\beta}} S_{\beta}^{+}S_{\alpha}^{-}\right) \, ,  \notag
\end{align}
where the definition of the spin operators $S_{\alpha}^{\pm}$ depends on the lattice site as 
described by Eqs.~(\ref{sb}) and~(\ref{sr}). The operators $S_{\alpha}^{\pm}$ obey Fermi anticommutation relations on the same site,
\begin{align}
 S_{\alpha}^{+}S_{\alpha}^{-}+ S_{\alpha}^{-}S_{\alpha}^{+} = \mathds{1}\,, \quad S_{\alpha}^{+}S_{\alpha}^{+}= S_{\alpha}^{-}S_{\alpha}^{-}=0\,,
 \label{fcom}
\end{align}
and Bose commutation relations between different sites,
\begin{align}
 \left[S_{\alpha}^{-},S_{\beta}^{+}\right]= \left[S_{\alpha}^{+},S_{\beta}^{+}\right]= \left[S_{\alpha}^{-},S_{\beta}^{-}\right] = 0\,,\quad \alpha\not=\beta\,.
 \label{bcom}
\end{align}
It follows that the raising and lowering operators $S_{\alpha}^{+}$ and $S_{\alpha}^{-}$ are equivalent to hard-core bosonic creation and annihilation operators 
$a_{\alpha}^{\dagger}$ and $a_{\alpha}$, respectively. The Hamiltonian in Eq.~(\ref{effham}) describes the hopping dynamics of these hard-core bosons on 
the two coupled sub-lattices $\mc{A}$ and $\mc{B}$. 
An example for the dipole-dipole coupling strengths in rubidium atoms and the  magnitude of the level shifts required for realizing 
the effective Hamiltonian in Eq.~(\ref{effham}) is provided in Appendix~\ref{appendixB}. Here 
we outline two physical implementations of the level shifts $\hat{L}_{\alpha}[l_j]$ in Eq.~(\ref{h0}).  
First, we consider linear Zeeman shifts induced by an external magnetic field $B_{\alpha}$ in $z$ direction, 
\begin{align}
\hat{L}_{\alpha}[l_{j}] = \frac{g[l_j]}{\hbar} \mu_B B_{\alpha} \hat{J}_z[l_j]\, , 
\label{ShiftOp}
\end{align}
where $\mu_B$ is the Bohr magneton,  $\hat{J}_z[l_j]$ is the $z$ component of the angular momentum operator restricted to the multiplet $l_j$, 
and $g[l_j]$ is the Land\'{e} g-factor, 
\begin{equation}
g[l_j] = \frac{3}{2} + \frac{3/4-l(l+1)}{2j(j+1)} \, . 
\end{equation}
Since $g[s_{1/2}]=2$ and $g[p_{3/2}]=4/3$, the magnitude of the Zeeman shifts is different 
for the $s_{1/2}$ and $p_{3/2}$ manifolds, respectively. 
We assume that atoms in lattices $\mc{B}$ and $\mc{R}$ experience different magnetic field strengths, 
\begin{align}
 B_{\alpha} = 
 \left\{
 \begin{array}{ll}
  B_{\mc{B}},&\alpha\in\mc{B}\,, \\[0.2cm]
  B_{\mc{R}},&\alpha\in\mc{R} \, ,
 \end{array}
 \right.
\end{align}
where $B_{\mc{B}}\not=B_{\mc{R}}$. Exact resonance $\Delta=0$ between the two sub-lattices  can be achieved for $B_{\mc{B}}=-5 B_{\mc{R}}/3$, and 
periodic magnetic fields  could be engineered by a regular array of micromagnets~\cite{ye:95,nogaret:10}. 

Second, the effective Hamiltonian in Eq.~(\ref{effham}) can  be realized  with a uniform magnetic field across all lattice sites 
and  static or AC Stark shifts that are different for the $\mc{B}$ and $\mc{R}$ lattices. For example, one could employ AC Stark 
shifts using a standing wave with periodicity $b$ such that all $\mc{B}$ and $\mc{R}$ atoms are located at the nodes and antinodes, respectively. 
Since the magnitude of the AC Stark shifts depends on $|m_j|$, a relative shift between the $\ket{p_{3/2}3/2}\leftrightarrow\ket{s_{1/2}1/2}$ 
and $\ket{p_{3/2}-1/2}\leftrightarrow\ket{s_{1/2}1/2}$ transitions can be induced such that the resonance condition $\Delta\approx 0$ holds.
\section{Results \label{results}}
%
\begin{figure}[t!]
\begin{center}
\includegraphics[width=7.5cm]{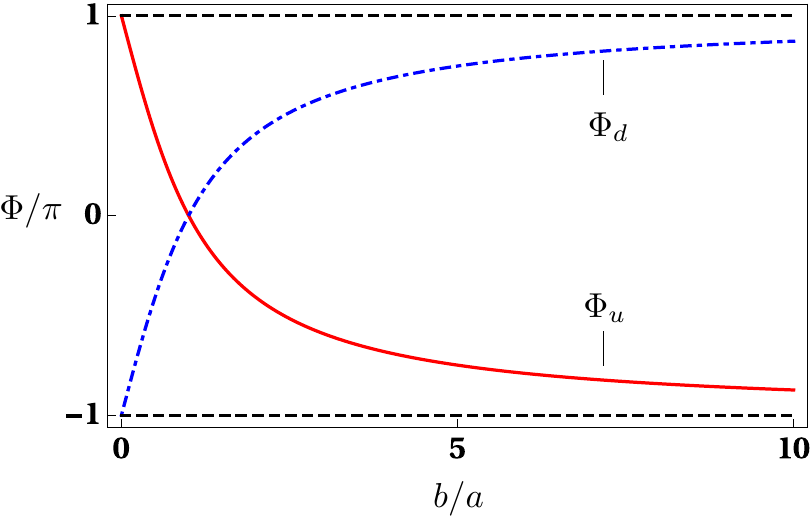}
\end{center}
\caption{\label{fig3}
(Color online)
Magnetic flux $\Phi$ enclosed in an elementary triangle of the lattice in Fig.~\ref{fig1}  as a function of $b/a$, where $b$ and 
$a$ are the lattice constants of the rectangular sub-lattices. 
$\Phi_u$ ($\Phi_d$) is the flux through the upward (downward) pointing  triangle $1\rightarrow2\rightarrow3\rightarrow1$ 
($2\rightarrow4\rightarrow3\rightarrow2$) in Fig.~\ref{fig1} 
and for the effective Hamiltonian in Eq.~(\ref{effham}) with only nearest neighbour interactions taken into account. 
}
\end{figure}
%
The  effective Hamiltonian in Eq.~(\ref{effham}) exhibits complex hopping amplitudes for exciton transitions between the 
$\mc{B}$ and $\mc{R}$ lattices which correspond to an artificial vector potential $\vec{A}$ according to the Peierls substitution~\cite{hofstadter:76}. 
This result can be understood as follows. Excitations in the $\mc{B}$ and $\mc{R}$ 
lattices couple to different dipole transitions with complex dipole moments $\vec{d}_{\mc{B}}$ and $\vec{d}_{\mc{R}}$, respectively. 
The two different transitions on sites $\mc{B}$ and $\mc{R}$ are tuned into resonance through external fields that 
break time-reversal symmetry. 
Since $\vec{d}_{\mc{B}}$ and $\vec{d}_{\mc{R}}$ have a well-defined relative phase, hopping between the 
two sub-lattices gives rise to a complex hopping amplitude that depends 
on the azimuthal angle of the relative position vector $\vec{R}_{\alpha}-\vec{R}_{\beta}$   between sites $\alpha$ and $\beta$, see Eq.~(\ref{phasedef}). 
The total magnetic flux $\Phi_u$ through the upward pointing triangle $1\rightarrow2\rightarrow3\rightarrow1$ 
is shown in Fig.~\ref{fig1}. For nearest neighbour interactions only, the total flux is determined by the sum of the phases along the edges of the triangle, 
\begin{align}
 \Phi_u & = 2(\phi_{32}-\phi_{13})+\pi\,. 
\end{align}
We find that the total flux is in general different from zero and can be adjusted by varying the 
lattice parameters. This is shown by the red solid line in Fig.~(\ref{fig3}), 
where $ \Phi_u$ is depicted as a function of ratio $b/a$. $ \Phi_u$ is different from zero except for $b/a=1$ 
and attains all possible values between $-\pi$ and $\pi$, which is the maximal range for the flux defined mod $2\pi$. 
Similarly, the total magnetic flux $\Phi_d$ through the downward pointing triangle $2\rightarrow4\rightarrow3\rightarrow2$ 
in Fig.~\ref{fig1} is given by 
\begin{align}
 \Phi_d & = 2(\phi_{34}-\phi_{23})+\pi\,,
\end{align}
and $\Phi_d$ is shown by the blue dot-dashed line in Fig.~\ref{fig3}. Since $\Phi_d+\Phi_u=0$ for all values $b/a$, 
the flux in neighbouring triangles has the same magnitude but the opposite sign, and hence the complex transition 
amplitudes in our system correspond to a staggered artificial magnetic field. This result is consistent with the 
assumed translational symmetry of the lattice, which requires that all magnetic fluxes within the unit cell must add up to zero. 

\begin{figure}[t!]
\begin{center}
\includegraphics[width=8.5cm]{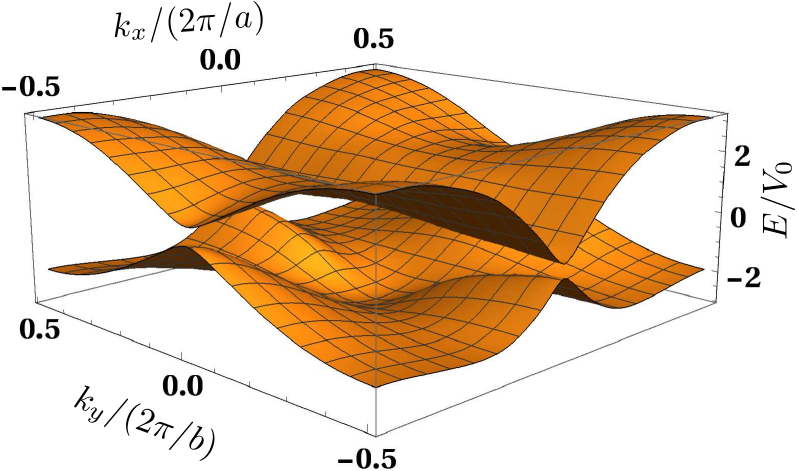}
\end{center}
\caption{\label{fig4}
(Color online)
Single-excitation band structure for $\Delta=0$ and $b/a=\sqrt{3}$. All hopping terms between sites within
a radius of $r_D=6 a$ are taken into account. The two bands are separated by a gap and the lower (upper) band has Chern number $C=1$ ($C=-1$). 
}
\end{figure}

Next we investigate the single-particle band structure of $H_{\text{eff}}$ using a rectangular unit cell containing two 
lattice sites as shown by the shaded area in Fig.~\ref{fig1}. It follows that the $k$-space  Hamiltonian $\mc{H}(\vec{k})$ 
is represented by a $2\times2$ matrix, where $\vec{k}$ describes a point in the first Brillouin zone of the reciprocal lattice (see Appendix~\ref{kspace}). 
We include all hopping terms between sites  separated by $R\le r_D$. 
Through a numerical study we find 
that $\mc{H}(\vec{k})$ describes the bulk properties of our system well for  $r_D \ge 6 a$ and $a\ge b/2$. 
The band structure for the special case of equilateral triangles as in Fig.~\ref{fig1} (i.e., $b/a=\sqrt{3}$) 
is shown in Fig.~\ref{fig4}. There are two separate bands and the band gap varies in size across the Brillouin zone. 
The gap is the smallest near the following points at the zone boundary, 
\begin{subequations}
  \label{kpoints}
\begin{align}
 & \vec{k}_1=(0, \pi/b)\, && \vec{k}_2=(0, -\pi/b) \, \\
 & \vec{k}_3=(\pi/a,0)\, && \vec{k}_4=(-\pi/a,0) \,. 
\end{align}
\end{subequations}
The magnitude of the band gap near these points is of the order of $ V_0/2$, where $V_0$ is 
defined in Eq.~(\ref{strength}). 
The  broken time-reversal symmetry in our system endows the band structure with 
non-trivial topological properties. We numerically  calculate the Chern number as described 
in~\cite{fukui:05} and find that the lower and upper bands have Chern numbers $C=1$ and 
$C=-1$, respectively. 
%
%

\begin{figure}[t!]
\begin{center}
\includegraphics[width=8cm]{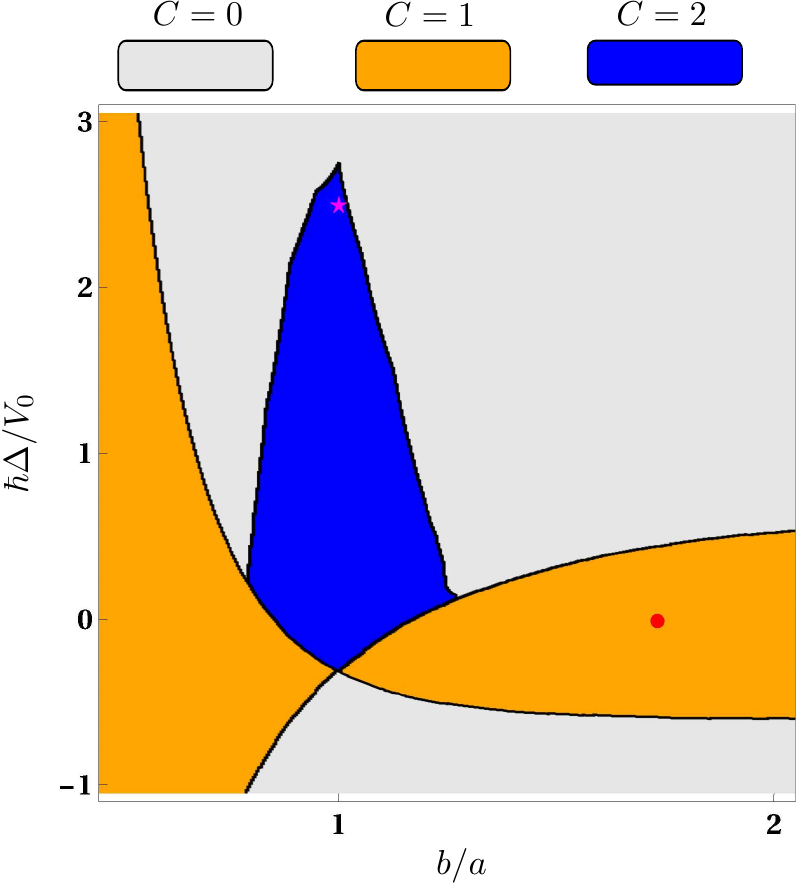}
\end{center}
\caption{\label{fig5}
(Color online)
Topological regimes for the lower band for $r_D=8$. The black lines indicate topological phase transitions where 
the Chern number of the lower band changes. The  red dot and magenta star correspond to the parameters of 
the band structures in Figs.~\ref{fig4} and~\ref{fig6}, respectively. 
}
\end{figure}

\begin{figure}[t!]
\begin{center}
\includegraphics[width=8cm]{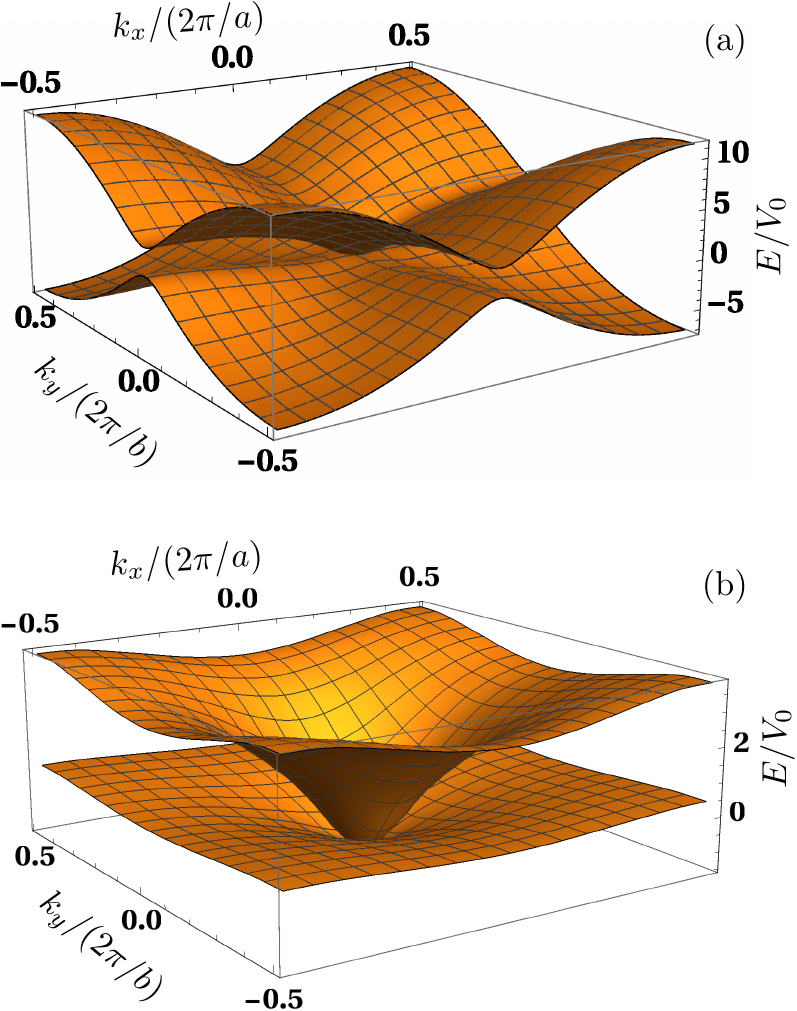}
\end{center}
\caption{\label{fig6}
(Color online)
Single-excitation band structure for $\hbar\Delta=2.5 V_0$ and $b/a=1$. All hopping terms between sites within
a radius of $r_D=6 a$ are taken into account. 
(a) Band structure corresponding to the effective Hamiltonian in Eq.~(\ref{effham}).
(b) Band structure for the same parameters as in (a) but without the Hamiltonian $H_{\mc{B}\mc{R}}=0$ coupling 
the two sub-lattices. The upper (lower) surface is the band structure for excitations on the $\mc{R}$ ($\mc{B}$) lattice. 
}
\end{figure}

%
Various  topological regimes can be realized in our system by adjusting the lattice parameters 
and the detuning between the excitations on the sub-lattices $\mc{B}$ and $\mc{A}$. 
This is illustrated in Fig.~\ref{fig5}, where we show the Chern number of the lower band 
as a function of the ratio $b/a$  and $\Delta$.  
First, we note that the phase diagram in Fig.~\ref{fig5} exhibits extended regions with non-zero Chern numbers 
that are robust with respect to small variations in $\Delta$ and $b/a$.
The solid lines in Fig.~\ref{fig5} indicate  topological phase transitions where 
the lower and upper bands touch in at least two of the $\vec{k}$ points in Eq.~(\ref{kpoints}) which 
then represent a Dirac point. 
The qualitative features of the $C=1$ region marked in orange in Fig.~\ref{fig5} can be understood by noting that non-zero Chern numbers 
require an efficient coupling between the sub-lattices $\mc{B}$ and $\mc{R}$. In particular, 
the dipole-dipole coupling needs to be larger or comparable to the detuning $\Delta$. For fixed lattice constant $a$, 
reducing $b/a$ corresponds to an increased  dipole-dipole coupling between the sub-lattices and  hence 
the region with $C=1$ broadens along the $\Delta$ axis for $b/a<1$. 
The narrowing of the $C=1$ region near $b/a=1$ can be understood from Fig.~\ref{fig3}. 
For nearest-neighbour interactions only,  the magnetic flux vanishes for $b/a=1$ and 
hence the corresponding bands would have Chern number $C=0$. Taking into account interactions beyond nearest neighbours 
 gives rise to modifications as shown in Fig.~\ref{fig5}. In particular, these interactions are responsible 
for  the blue wedged area with Chern number $C=2$. 
%
%
The single-particle band structure for the parameters corresponding to the magenta star inside the blue wedged area in Fig.~\ref{fig5} 
is shown in Fig.~\ref{fig6}(a). The lower and upper bands have Chern numbers $C=2$ and $C=-2$, respectively. The two bands are gapped, but 
in contrast to the parameters in Fig.~\ref{fig4} the gap is the smallest near the Brillouin zone center $\vec{k}=(0,0)$ where it is approximately given by $0.1 V_0$.

The asymmetry of the phase diagram in Fig.~\ref{fig5} with respect to the $\Delta=0$ axis can 
be traced back to the fact that the dipole-dipole interaction differs in strength for 
the $\mc{B}$ and $\mc{R}$ lattices. In order to illustrate this, we focus on the 
blue wedge with $C=2$ in Fig.~\ref{fig5} and show the band structures of the uncoupled, individual sub-lattices in Fig.~\ref{fig6}(b) for $\hbar\Delta=2.5 V_0$ 
and $b/a=1$. 
Both band structures are  convex surfaces with their minimum at $\vec{k}=0$, 
but the depth of the potential well is significantly larger for the upper band. The reason  is that   
the strength of the dipole-dipole interaction is three times stronger for the  
$\mc{R}$ lattice compared to the $\mc{B}$ lattice for $b/a=1$, see Eqs.~(\ref{hb}) and~(\ref{hr}). 
A necessary condition for non-trivial topological bands is that the two sub-lattices are efficiently 
coupled by the Hamiltonian $H_{\mc{B}\mc{R}}$ in  Eq.~(\ref{hbr}), which depends on the magnitude of the 
dipole-dipole interaction connecting the $\mc{B}$ and $\mc{R}$ lattices and the energy spacing between $\mc{B}$ and $\mc{R}$ 
excitations at each $\vec{k}$ point. 
As can be seen in Fig.~\ref{fig6}(b), the two surfaces  touch near
$\vec{k}\approx\vec{0}$, and hence the relatively weak next-nearest neighbour coupling in $H_{\mc{B}\mc{R}}$ 
can give rise to non-zero Chern numbers for $\hbar\Delta=2.5 V_0$ and $b/a=1$. 
On the other hand,  the distance between the two uncoupled bands 
increases quickly if  $\Delta$ is decreased from zero to negative values. This explains why 
$H_{\mc{B}\mc{R}}$ cannot induce a $C=2$ band for $\hbar\Delta\lesssim -0.3V_0$.
%
%

%
Finally we discuss the physical realization of our system and the observation of its topological features. 
The experimental realization of a one-dimensional chain of resonantly coupled Rydberg atoms has been reported in~\cite{barredo:15}. 
Here the excitation of all atoms to a Rydberg state is achieved within $\tau\approx 0.5\mu\text{s}$~\cite{barredo:15}. Note that this process is not hampered 
by the dipole blockade since the van der Waals shifts are small for the considered lattice constants $a$. 
For example, for Rubidium $ns$ states with $n=70$ and $a=20\mu\text{m}$, the van der Waals shift is $\Delta_{\text{vdW}}\approx 13\,\text{kHz}$~\cite{singer:05}, 
which is small compared with the Rabi frequency of the lasers  exciting the Rydberg state~\cite{barredo:15}. 
The time interval $\Delta T$ where  excitation hopping can take place is limited by the lifetime of the Rydberg states and the residual atomic motion. 
For atomic temperatures of the order of $10\mu\text{K}$, motional 
effects are negligible for $\Delta T \approx 10\mu\text{s}$~\cite{barredo:15}. This  is typically much smaller than the Rydberg state lifetime and large 
compared with the inverse hopping amplitude such that many coherent hops can take place, see Appendix~\ref{appendixB}.
Note that these considerations also show that autoionisation processes due to Rydberg atom collisions can be neglected~\cite{amthor:07,kiffner:16b} 
since the initial positions of the atoms in the lattice change only very slightly during $\Delta T$. 
Recently, tremendous experimental progress towards the extension of the experiment in~\cite{barredo:15} to two dimensions and 
arbitrary lattice geometries has been made~\cite{nogrette:14,labuhn:16}. In particular, 
it is now possible to create arbitrary lattice structures where each site is filled with exactly one atom~\cite{barredo:16}.  
It follows that our system can be realized with a combination of state-of-the-art experimental techniques. 

%
\begin{figure}[t!]
\begin{center}
\includegraphics[width=8cm]{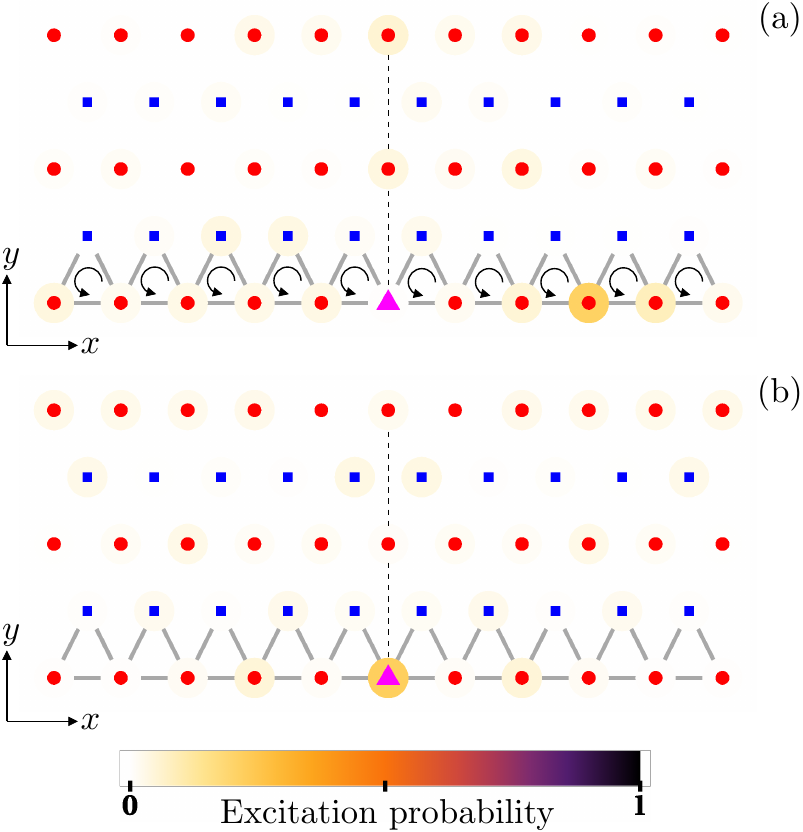}
\end{center}
\caption{\label{fig7}
(Color online)
Quantum dynamics of a single excitation on a lattice with 53 atoms, $b/a=2$, $\Delta=0$ and $r_D=6 a$.  
At time $t=0$, only the atom in the bottom row indicated by a magenta triangle is excited. 
The population of each lattice site at time $t = 4 \hbar/V_0$ is indicated by the color of the halo around each site. 
The dashed line is used as a guide to the eye (see text). 
(a) Quantum dynamics according to the Hamiltonian $H_{\text{eff}}$ in Eq.~(\ref{effham}). The magnetic flux through the 
indicated triangular plaquettes is negative and thus favours counter-clockwise motion of the excitation.
(b) Same as in (a), but with all phases $\phi_{\alpha\beta}$  in Eq.~(\ref{effham}) set to zero.
}
\end{figure}
%
A direct signature of the artificial magnetic fields associated with the complex hopping amplitudes in Eq.~(\ref{effham}) can be obtained by 
investigating the quantum dynamics of a single excitation as shown in Fig.~\ref{fig7}. We consider a  lattice with 
53 sites where only the site in the middle of the lower edge is excited at time $t=0$. 
The excitation probability of the lattice sites at a later time is shown in Figs.~\ref{fig7}(a) and~(b), where Fig.~\ref{fig7}(a) 
the dynamics according  to the effective Hamiltonian in Eq.~(\ref{effham}). We find that the largest excitation probabilities can be found along the 
lower edge and to the right of the initially excited site. 
Figure~\ref{fig7}(b) was generated by setting all 
phases $\phi_{\alpha\beta}$ in Eq.~(\ref{effham}) to zero. In this case, the distribution of excitation probabilities is symmetric with respect to 
the dashed line. The latter result is expected since the magnitude of the hopping amplitudes does only depend on the distance between two sites. 
It follows that the marked asymmetry in Fig.~\ref{fig7}(a) is a direct consequence of the complex hopping amplitudes and the associated artificial magnetic field. 
More specifically,  the magnetic flux through the upward pointing triangles 
shown in Fig.~\ref{fig7} and for the considered parameters is negative, see Fig.~\ref{fig3}. The force associated with the artificial magnetic field 
thus favours an anti-clockwise motion around each triangular plaquette. This explains why the propagation moves along the edge in an anti-clockwise direction. 
Note that this asymmetry develops within a few hopping events such that the residual motion of the atoms hosting these excitations can be neglected. 
Our results are also consistent with the fact that a semi-infinite version of our lattice exhibits chiral edge states for non-zero 
Chern numbers according to the bulk-edge correspondence~\cite{hatsugai:93,qi:06}. 
The Chern number of the individual bands can be determined by observing the motional drifts due to the non-zero Berry curvature in each band. 
To this end, the excitations need to be selectively prepared in either the upper or lower band. This can be achieved in different ways. 
First, one could prepare an excitation in one of the sub-lattices with a large detuning $\Delta$ such 
that the $\mc{R}$ and $\mc{B}$ lattices are uncoupled. This is followed by an adiabatic reduction of $|\Delta|$ in order to adjust the required parameter regime. 
Second, one could prepare all atoms in the $\ket{ns_{1/2}1/2}$ state and apply a weak microwave field such that only 
a  single $\vec{k}$ mode is resonantly excited. 
Efficient methods to extract the local Berry curvature from motional drifts are described in~\cite{price:12} 
and require an external force acting on the particle. In our setup, this could be realized by making the detuning $\Delta$ position-dependent through magnetic field gradients 
along a certain direction. 
\section{Summary \label{summary}}
We have shown that the resonant dipole-dipole interaction between Rydberg atoms allows one 
to engineer effective spin-1/2 models where the spin excitations experience  a  staggered magnetic field 
in a triangular lattice. A necessary condition for engineering artificial magnetic fields is that 
time-reversal symmetry of the system is broken. In our system, this is achieved by  external fields 
shifting the Zeeman sublevels of the considered $ns_{1/2}$ and $np_{3/2}$ manifolds. In this way we ensure 
that the spin excitation couples to different dipole transitions on the $\mc{B}$ and $\mc{R}$ lattices with 
dipole moments $\vec{d}_{\mc{B}}$ and $\vec{d}_{\mc{R}}$, respectively. These dipole moments have a well-defined
relative phase which is different from zero. Since $\vec{d}_{\mc{B}}$ and $\vec{d}_{\mc{R}}$ are orthogonal, 
we find that the phase of the hopping amplitude is determined by the azimuthal angle associated with the relative 
position of the two sites connected by the hop. 

We find that the  magnitude of the magnetic flux through an elementary triangular plaquette can be controlled by 
changing the ratio $b/a$ of the rectangular sub-lattices. In addition, the 
staggered magnetic field endows the single-particle band structure with non-trivial Chern numbers.  
The Chern number of the lower band can be adjusted between $C=0,\,1$ and $2$ and its value depends 
on the lattice parameters and the detuning $\Delta$ between the $\mc{B}$ and $\mc{R}$ lattices. 
The quantum simulation of the dynamics of a single excitation shows that an excitation placed at an edge of 
the lattice will propagate along the edge in a specific direction. This effect is  
a direct consequence of  the artificial magnetic field. 
The topological features of the bands  can be explored by monitoring the 
deflection of the exciton motion due to the non-zero Berry curvature in either the lower or upper band. 
An intriguing prospect for future studies is the investigation of quantum many-body states. Here the 
hard-core interaction between the particles is expected to modify the single-particle picture considerably, 
and the interplay of strong interactions and complex hopping amplitudes may give rise to exotic quantum phases like fractional Chern insulators. 
\begin{acknowledgements}
MK  thanks the National Research Foundation and the Ministry of Education of
Singapore for support and Tilman Esslinger for helpful discussions.
The authors would like to acknowledge the use of the University of Oxford Advanced Research Computing (ARC) facility in 
carrying out this work (http://dx.doi.org/10.5281/zenodo.22558).
\end{acknowledgements}
\appendix
\section{Dipole matrix elements\label{appendixA}}
We evaluate the matrix elements of the electric-dipole-moment operator $\mf{\hat{d}}$ of an individual atom 
via the Wigner-Eckert theorem~\cite{walker:08,edmonds:amq} and find 
\begin{align}
\bra{nl'_{j'}m'}\mf{\hat{d}}\ket{nl_{j}m} & = \mc{D} \sum_{q=-1}^1 
C_{jm1q}^{j' m'} \vec{\mf{\epsilon}}_q \, , \label{we1}
\end{align}
where $C_{jm1q}^{j\ensuremath{'}m\ensuremath{'}}$ are Clebsch-Gordan
coefficients and the spherical unit vectors $\vec{\mf{\epsilon}}_q$ in Eq.~(\ref{we1}) are defined as 
\begin{eqnarray}
\vec{\epsilon}_{1}=-\frac{\vec{e}_x-i \vec{e}_y}{\sqrt{2}},\quad
\vec{\epsilon}_{0}=\vec{e}_z,\quad
\vec{\epsilon}_{-1}=\frac{\vec{e}_x+i \vec{e}_y}{\sqrt{2}} \,.
\label{vecs}
\end{eqnarray}
The reduced dipole matrix element is~\cite{walker:08,edmonds:amq} 
\begin{align}
\mc{D} = & (-1)^{\mathrm{j+l'-1/2}}\sqrt{2j+1}\sqrt{2l+1} \notag \\
&
\left \lbrace 
\begin{array}{ccc}
l' & l & 1 \\
j & j' & 1/2
\end{array} 
\right \rbrace  
C_{10l0}^{l'0} e \bra{n' l'}r\ket{nl} \, ,
\label{D}
\end{align}
where the $3\times 2$ matrix in curly braces is the Wigner $6-j$ symbol,   $e$ is the elementary charge and
$\bra{n\ensuremath{'}l\ensuremath{'}}r\ket{nl}$ is a radial matrix element. 
\section{Rubidium parameters\label{appendixB}}
Here we calculate the strength of the dipole-dipole interaction for rubidium atoms 
and estimate the magnitude of the level shifts required for realizing our model. 
For $ns_{1/2}\leftrightarrow np_{3/2}$ transitions in rubidium with principal quantum number $n=70$, 
the reduced dipole moment $\mc{D}$ in Eq.~(\ref{D}) is given by
\begin{align}
 \mc{D} \approx 2909 e a_0\,,
\end{align}
where $e$ is the elementary charge and $a_0$ is the Bohr radius. It follows that 
the strength of the dipole-dipole coupling $V_0$ in Eq.~(\ref{strength}) for 
$a=20\mu\text{m}$ is 
\begin{align}
 V_0/\hbar \approx 2\pi\times 1.03 \,\text{MHz}\,.
\end{align}
The lifetime of the $ns_{1/2}$ and $np_{3/2}$ states at temperature $T=300\text{K}$ and for $n=70$ is $T_s\approx 151.6\mu\text{s}$ and $T_p\approx191.3\mu\text{s}$, respectively~\cite{beterov:09}. 
Note that these values take into account the lifetime reduction due to  blackbody radiation. 
The hopping rates vary with the lattice parameters but are typically of the order of $V_0$. It follows that 
in principle many coherent hopping events can be observed before losses due to spontaneous emission set in. 
This finding is consistent with the experimental observations in~\cite{barredo:15}. Note that the magnitude of 
$V_0$ can be  increased by reducing the size of the lattice constant $a$ or by increasing $n$. 
Next we discuss the requirements for reducing the general Hamiltonian in Eq.~(\ref{fullham}) to our model in Eq.~(\ref{effham}). 
First, we note that the level shifts induced between Zeeman substates must be large compared to $V_0$ and hence of the order of $10\,\text{MHz}$. 
Shifts of this magnitude can be realized with weak magnetic fields~\cite{steck:dl} or AC stark shifts~\cite{barredo:15}. 
Furthermore, the fine structure splitting between the $ns_{1/2}$ and $np_{3/2}$ manifolds is $\Delta E_{\text{FS}}\approx 2\pi\times 10.8\text{GHz}$~\cite{li:03}, 
which is much larger than 
$V_0$ and hence it is justified to neglect off-resonant  terms in Eq.~(\ref{vdd}). 
Finally, we note that the energy difference between the  $np_{3/2}$ manifold and the nearby $np_{1/2}$ manifold is approximately $285\,\text{MHz}$~\cite{li:03}, 
which is also much larger than $V_0$. It follows that the $np_{1/2}$ states can be safely neglected. 
\section{$k$-space Hamiltonian \label{kspace}}
The $k$-space Hamiltonian can be obtained by considering the single-excitation subspace $\mc{E}_1$ spanned by the 
basis states 
\begin{align}
 \ket{\alpha}= S_{\alpha}^+\ket{0}\,,
\end{align}
where $\ket{\alpha}$ denotes one $p$ excitation at site $\alpha$ and $\ket{0}$ is the ``vacuum'' state with 
zero excitations, i.e., the atoms at all lattice sites are in state $\ket{s_{1/2}1/2}$. 
In order to solve the eigenvalue equation
\begin{align}
 H_{\text{eff}}\ket{\psi} = E \ket{\psi} 
 \label{eval}
\end{align}
with  $\ket{\psi}\in\mc{E}_1$, we describe the lattice in Fig.~\ref{fig1} by a rectangular Bravais lattice with a two-atomic 
basis. More specifically, the direct lattice points are given by the $\mc{R}$ atoms such that the basis is comprised of 
one $\mc{R}$ atom at $\vec{0}$ and one $\mc{B}$ atom at $(\vec{a}+\vec{b})/2$. 
According to Bloch's theorem~\cite{ashcroft:sp}, we can solve Eq.~(\ref{eval}) with the Ansatz
\begin{align}
 \ket{\psi}= \sum\limits_{\alpha} u_{\alpha} \ket{\alpha}\,,
 \label{ansatz}
\end{align}
where the coefficients $u_{\alpha}$ can be written as 
\begin{align}
 u_{\alpha} = 
 \left\{
 \begin{array}{ll}
  \psi_{\mc{R}}e^{\imag \vec{k}\cdot \vec{U}(\alpha)} ,& \alpha\in\mc{R}\,, \\[0.2cm]
  \psi_{\mc{B}}e^{\imag \vec{k}\cdot \vec{U}(\alpha)} ,& \alpha\in\mc{B} \, ,
 \end{array}
 \right.
 \label{ansatzD}
\end{align}
and $\vec{k}$ is a point in the first Brillouin zone of the direct lattice.
The vector $\vec{U}(\alpha)$ in Eq.~(\ref{ansatzD}) is the Bravais lattice point associated with site $\alpha$, 
\begin{align}
\vec{U}(\alpha) = 
 \left\{
 \begin{array}{ll}
  \vec{R}_{\alpha} ,& \alpha\in\mc{R}\,, \\[0.2cm]
  \vec{R}_{\alpha}-(\vec{a}+\vec{b})/2 ,& \alpha\in\mc{B} \, .
 \end{array}
 \right.
\end{align}
With Eqs.~(\ref{ansatz}) and~(\ref{ansatzD}), Eq.~(\ref{eval}) can be reduced to the following 
matrix equation for the amplitudes $\psi_{\mc{R}}$ and $\psi_{\mc{B}}$, 
\begin{align}
 \mc{H}(\vec{k})
\left(
 \begin{array}{l}
  \psi_{\mc{R}}  \\[0.2cm]
  \psi_{\mc{B}} 
 \end{array}
 \right) = 
 E
 \left(
 \begin{array}{l}
  \psi_{\mc{R}}  \\[0.2cm]
  \psi_{\mc{B}} 
 \end{array}
 \right) \,,
\end{align}
where the $2\times2$ matrix  $\mc{H}(\vec{k})$ is the $k$-space Hamiltonian. 
We find $\mc{H}(\vec{k})$ using the software package MATHEMATICA~\cite{MM} for each set of lattice parameters $a$ and $b$. 
In general, the resulting expressions are too complicated 
to display here. In the special case of   nearest-neighbour interactions only, we find 
\begin{subequations}
\begin{align}
[\mc{H}(\vec{k})]_{11} & = -V_0 \cos(\vec{k}\cdot\vec{a})+ \hbar\Delta \,, \\
 [\mc{H}(\vec{k})]_{12} & =\frac{V_0 \sqrt{48}}{[1+(b/a)^2]^{3/2}}   \left(e^{- \imag(\vec{k}\cdot\vec{a}+\vec{k}\cdot\vec{b})}e^{-2\imag\alpha}+e^{-2\imag\alpha}
 \right. \notag \\
 &\qquad \quad \left. +e^{-\imag\vec{k}\cdot\vec{a}}e^{2\imag\alpha} +e^{ -\imag\vec{k}\cdot\vec{b}}e^{2\imag\alpha}\right) \,, \\
 [\mc{H}(\vec{k})]_{22} & = -\frac{1}{3} V_0 \cos(\vec{k}\cdot\vec{a})\,,
 \end{align}
\end{subequations}
where $\cos(\alpha)=1/[1+(b/a)^2]^{1/2}$ and $[\mc{H}(\vec{k})]_{21}=[\mc{H}(\vec{k})]_{12}^*$. 
\end{document}